\documentstyle[aps,12pt]{revtex}
\begin{document}

\draft
\title{Effects of disorder and magnetic field in frustrated magnets}
\vspace{2cm}

\author{Adauto J. F. de Souza}
\address{Departamento de F\'{\i}sica e Matem\'atica, Universidade
Federal Rural de Pernambuco, 52171-030 Recife PE, Brazil \\
Departamento de F\'{\i}sica, Universidade Federal
de Pernambuco, 50670-901 Recife - PE, Brazil}
\vspace{1cm}
\author{F.G. Brady Moreira and J.R.L. de Almeida}
\address{Departamento de F\'{\i}sica, Universidade Federal de
Pernambuco,
50670-901 Recife - PE, Brazil}
\maketitle

\vspace{2cm}

\begin{abstract}

In this work a site diluted antiferromagntic Ising model in the FCC
lattice is studied by Monte Carlo simulation. At low temperatures, we find that as the external field
 is increased the transition from the antiferromagnetic phase to the superantiferromagnetic one occurs
through an intermediate phase which is not present in the undiluted system. This new phase ordering
 has three distinct values for the sublattice magnetizations corresponding to one of the phases found
in a recent mean field calculation  thus suggesting that in strongly frustrated systems many novel
spins ordering may arises as found experimentally, for instance, in some pyrochlores.


 \end{abstract}

\vspace{2cm}

Short title: {\bf Effects of disorder in frustrated magnets}

 \vspace{2cm}

e-mail : almeida@df.ufpe.br

\newpage

\section{Introduction}

 Recently a lot of work has been done on frustrated systems with
 or without disorder in order to clarify the physical properties of
systems exhibiting glassy behavior\cite{ref1}, i.e.,  systems typically with
many metastable states. Among classical fully frustrated systems, one
of the earliest to be studied was the nearest-neighbor Ising antiferromagnet
 on a face-centered cubic (FCC) lattice. The pure FCC Ising antiferromagnet
 with nearest –neighbor interactions and in the presence of a uniform magnetic
field H, exhibits a variety of stable phases induced by the external field.
The field-temperature $(H,T)$ phase diagram, which follows from Monte
 Carlo simulations\cite{grillon}presents three phases: antiferromagnetic (AF),
superantiferromagnetic (SAF), and paramagnetic (PM). This picture
is supported by several theoretical approaches and it is well known that
the low temperature ordered phases are separated from the paramagnetic
phase by two lines of first-order phase transitions. A mean field
 calculation on the random version of the FCC Ising antiferromagnet
in a field shows  an even richer scenario\cite{jairo}. Beside the phases present
 in the undiluted case, it was found that the addition of disorder may
induce another ordered phase where two sublattices have equal
magnetizations which is distinct from the others  not equal two
 sublattice magnetizations. Furthermore, the disordered system presents
glassy behavior at low temperatures\cite{jairo,grest}. In this letter, we address the
 issue whether the mean field results are stable against fluctuations in
finite dimensions by doing Monte Carlo simulations on the site diluted
 FCC Ising antiferromagnet and our results seems to vindicate the
qualitative mean field results.

\section{The model, simulation and discussion}
  
We consider an Ising model with short-range interaction  on a regular
 lattice (FCC) with Hamiltonian given by
\begin{equation}
{\cal H} = \sum_{(ij)}J_{ij}\epsilon_i \epsilon_j  \sigma_i \sigma_j - H \sum_i \sigma_i,
\end{equation}
where $J_{ij}= J > 0$ denotes the antiferromagnetic interaction between nearest neighbors spins,
$ \epsilon_i=0$  or $1$ denotes the site occupation variable (probability $p$ of being occupied),
$\sigma_i = \pm 1$ denotes the spin on the sites of the FCC lattice and H is the external field. The
 simulations were performed on lattices of sizes $4L^3 (L=20,40)$ with periodic
 boundary conditions, and for concentrations of magnetic ions between $p=0.70$
 and  $p=1.00$. We used both the heat-bath and the demon (microcanonical)
algorithms. Most of our runs consisted of $4\times 10^3$ configurations, $5$ to $10$ MCS
 apart, and the quoted values for the physical observables came from an
 average over several independent runs on different disorder realizations.
Typically, $\sim 10^4$ MCS were discarded before we start measurements. Several
variants of walking through the lattice were considered and, particularly for
 low temperatures, the results were rather sensitive to this choice. We
 employed the single demon algorithm to accomplish the microcanonical
simulation\cite{ref5}. It is worth noticing that in a microcanonical simulation the
 energy is the control parameter, whereas the temperature is measured during
the simulation. Most important, within this framework it is possible to probe
 stable as well as both metastable and unstable states, and a first-order phase
 transition manifests itself through a S-shaped curve in the temperature-energy
 plane\cite{grillon,ref5}.

We have performed Monte Carlo simulations along of some lines of
the $(H,T)$ phase diagram, and for several values of $p$. Here, we will present results
for $p=0.95$ only. Figure 1 shows the field dependence of the sublattice magnetizations
 for $k_B T/J=1.3$ and magnetic concentration $p=0.95$. At this temperature and
 concentration one has the same behavior as in the pure case, namely, the AF
and SAF phases are separated by a paramagnetic phase at intermediate fields.
Our microcanonical simulation also indicates that both transitions are of first order
 as in references\cite{grillon}.

Lowering $T$, the sublattice magnetizations exhibit a quite different
behavior, as shown in figure 2 for  $k_B T/J=1.0$ and the same value of  $p$.
 One can observe the suppression of the paramagnetic phase between the
AF and SAF phases; in this region the sublattice magnetizations have a field
 dependence not observed in the pure system. At lower temperatures this
behavior is enhanced as ones sees in figure 3 for $k_BT/J=0.90$. For some
 points in the $(H,T)$ space, where the sublattice magnetizations show a
very slow relaxations towards their equilibrium values (within the time
scale of our simulations), we needed to make runs as long as $10^5$ to $10^6$  MCS.

In figure 4 we plot the sublattice magnetizations as a function of
 the temperature for $H/J=3.6$ and $p=0.95$. For this value of $H$ and high
enough T the ordering is paramagnetic. Lowering T the system passes through
a SAF phase before exhibiting the unexpected behavior as predicted in mean field
calculations\cite{jairo}. Again, we cannot discard that we are observing long-lived
metastable states at low temperatures from the simulations alone but at least
 some (if not all) qualitative features of mean field calculations are present for
 finite dimensions.

In summary, the FCC Ising antiferromagnet in an external magnetic
 field seems to change drastically its low temperature behavior when a small
amount of disorder is added. It is found that a typical order-by-disorder
 effect\cite{ref6} occurs upon dilution. The new sublattice magnetization ordering
that follows from our simulation corroborates the suggestion of a mean field
 calculation carried out for frustrated random magnetic systems with many
sublattices\cite{jairo}. The huge relaxation times which we have observed for some
values of field and low temperatures indicate that the diluted system might
 exhibit a glassy phase even for low dilution as suggested by mean field\cite{jairo}
and found experimentally in certain strongly frustrated pyrochlores systems
 (see the work of Bellier et al in \cite{ref1}and references therein ) which also have
a four sublattice structure. A more detailed account of our results should be
 presented elsewhere.

\vspace{1cm}
 Acknowledgments: the authors are grateful to CNPq
and FINEP for financial support.

\newpage
\centerline{\bf Figure Captions}
 \vskip 2em
\begin{description}

\item{Fig.~1} Sublattice magnetizations (indicated by squares, circles, up and down triangles)
 as function of the magnetic field for $k_BT / J = 1.3$ and $p=0.95$. Notice the presence of the
 paramagnetic (PM) phase between the antiferromagnetic (AFM) and superantiferromagnetic (SAF) phases.

\item{Fig.~2} Sublattice magnetizations as function of the magnetic field for $k_BT / J = 1.0$ and $p=0.95$.
The paramagnetic phase occurs only at high fields (not shown). The arrows are a guide to the eye
 showing the boundaries of the intermediate phase.

\item{Fig.~3} Same as in figure 2,  but for $k_BT / J = 0.90$. Notice the increase of the new
 intermediate phase between the antiferromagnetic (AFM) and
 superantiferromagnetic (SAF) phases as T decreases.

\item{Fig.~4} Sublattice magnetizations as function of temperature for $H / J = 3.6$ and $p=0.954$.
 This graph displays clearly the branching of the sublattice magnetizations in distincts
values as $T$ decreases (phases  PM  SAF  INTERMEDIATE) .

\end{description}

\end{document}